\documentclass[prb,nofootinbib,reprint,showpacs,superscriptaddress,amsmath,amssymb]{revtex4-1}

\usepackage{subfigure}
\usepackage{color}
\usepackage{graphicx}
\usepackage{dcolumn}
\usepackage{bm}
\usepackage{multirow}
\usepackage{hyperref}
\usepackage{natbib}

\usepackage{pstricks}

\DeclareMathOperator{\Tr}{Tr}

\begin{document}

\title{Dynamic universality class of Model C from the functional renormalization group}

\author{D.~Mesterh\'azy}
\email{mesterhazy@thphys.uni-heidelberg.de}
\author{J.H.~Stockemer}
\email{stockemer@thphys.uni-heidelberg.de}
\author{L.F.~Palhares}
\email{palhares@thphys.uni-heidelberg.de}
\author{J.~Berges}
\email{berges@thphys.uni-heidelberg.de}
\affiliation{
  Institut f\"ur Theoretische Physik, Universit\"at Heidelberg\\ 
  Philosophenweg 16, 69120 Heidelberg, Germany
}

\date{\today}

\begin{abstract}
We establish new scaling properties for the universality class of Model C, which describes relaxational critical dynamics of a nonconserved order parameter coupled to a conserved scalar density. We find an \emph{anomalous diffusion phase}, which satisfies weak dynamic scaling while the conserved density diffuses only asymptotically. The properties of the phase diagram for the dynamic critical behavior include a significantly extended \emph{weak} scaling region, together with a \emph{strong} and a \emph{decoupled scaling} regime. These calculations are done directly in $2 \leq d \leq 4$ space dimensions within the framework of the nonperturbative functional renormalization group. The scaling exponents characterizing the different phases are determined along with subleading indices featuring the stability properties.  
\end{abstract}
 
%dynamic critical behavior, phase transitions in renormalization-group theory, renormalization in field theory

\maketitle

Dynamic properties such as transport coefficients or relaxation rates play a crucial role for a wide variety of physical systems. Irrespective of the details of the underlying microscopic dynamics, they can be grouped into dynamic universality classes close to a critical point. Following the standard classification scheme~\cite{Hohenberg:1977ym}, the universality class of Model C is characterized in terms of an $N$-component order parameter with relaxational dynamics coupled to a diffusive field. Apart from being a model for the coupling of the energy density for Ising-like systems close to criticality, it is applied to the critical dynamics of mobile impurities \cite{Krey:1976aa,*Grinstein:1977aa}, structural phase transitions \cite{Halperin:1976aa,*Bausch:1978,*Prelovsek:1980aa}, long-wavelength fluctuations near the QCD critical point \cite{Berdnikov:1999ph,*Son:2004iv}, and out-of-equilibrium dynamics \cite{Calabrese:2002rk,*Akkineni:2004}.

Despite its importance and a long history of discussions~\cite{Halperin:1974zz,*Halperin:1976zz,Brezin:1975tq,Murata:1976,DeDominicis:1977fw,Folk:2003aa,*Folk:2004aa}, parts of the phase diagram for the dynamic critical behavior of Model C are still controversial. The reason for this uncertainty is that the physics is nonperturbative and only a few theoretical approaches apply. Previous calculations have mainly relied on the $\epsilon$ expansion in $d = 4 - \epsilon$ dimensions, while direct numerical simulations~\cite{Berges:2009jz} still represent an exception. While the existence of the so-called \emph{weak}, \emph{strong}, and \emph{decoupled scaling} regions is undebated, there have been conflicting claims on important quantitative properties and even on the possible existence of another distinctive region in the phase diagram of Model C as a function of $N$ and $d$. Earlier results~\cite{Halperin:1974zz,*Halperin:1976zz,Brezin:1975tq,Murata:1976,DeDominicis:1977fw} found evidence for such a region, however, it was unclear whether it persists to higher orders in the $\epsilon$ expansion. Other results to second order showed that for the ratio of kinetic coefficients an essential singularity occurs in this region \cite{Murata:1976}. It was speculated that this property might even restore critical behavior with a dynamic scaling exponent identical to the strong scaling exponent. In more recent work \cite{Folk:2003aa,*Folk:2004aa} this region was discarded as an artifact of the $\epsilon$ expansion, which was argued to break down for $2 < N < 4$ close to $d = 4$.

In this paper we compute the $(N,d)$ phase diagram for the dynamic critical behavior of Model C using the functional renormalization group, which is a nonperturbative approach that does not rely on the $\epsilon$ expansion. We establish an \emph{anomalous diffusion phase} with new scaling properties: It satisfies weak scaling for $2 < N < 4$ close to $d = 4$, however, the conserved density diffuses only on asymptotic times.
The properties of the phase diagram include a significantly extended weak scaling region to the whole range of $2 \leq d \leq 4$ for small $N$. These results show that the earlier proposed scaling solution at $N=0$ and $d=4$ belongs to a continuous phase boundary connecting to $N=4$ between the weak and strong scaling regimes. 

The functional renormalization group has been successfully applied to the calculation of static equilibrium critical properties \cite{Berges:2000ew}, to the dynamic critical scaling for purely relaxational models \cite{Canet:2006xu}, to field theories driven to a nonequilibrium steady state \cite{Canet:2009vz,*Canet:2011wf,Sieberer:2013}, as well as to stationary transport solutions described by nonthermal fixed points \cite{Berges:2008sr,*Berges:2012ty}. This paper presents the first determination of the dynamic critical properties of relaxational models in the framework of the functional renormalization group including the dynamics of conserved quantities. Such an analysis can be extended to also investigate other dynamic universality classes, or even to connect the dynamic low-energy properties with the microscopic physics of relativistic theories such as QCD.

The effective dynamics for Model C is given by \cite{Hohenberg:1977ym}
\begin{eqnarray}
\frac{\partial}{\partial t} \varphi_{a}(t) &=& - \Omega\, \frac{\delta \mathcal{H}[\varphi,\varepsilon]}{\delta \varphi_{a}(x,t)} + \eta_{a}(x,t) ~, 
\label{Eq:RelaxationalDynamics} \\
\frac{\partial}{\partial t} \varepsilon(t) &=& \Omega_{\varepsilon} \nabla^{2} \frac{\delta \mathcal{H}[\varphi,\varepsilon]}{\delta \varepsilon(x,t)} + \zeta(x,t) ~ ,
\label{Eq:DiffusiveDynamics}
\end{eqnarray}
with the stochastic driving terms $\eta_{a}$ and $\zeta$ and the equilibrium functional
\begin{equation}
\mathcal{H} =\! \int\! d^{d}x \left\{ \frac{1}{2} \left(\nabla \varphi\right)^{2} + \frac{\bar{r}}{2} \varphi^{2} + \frac{\bar{g}}{4!} \left(\varphi^{2}\right)^{2}
+\frac{1}{2} \varepsilon^{2} + \frac{\bar{\gamma}}{2} \varepsilon \varphi^{2} \right\} \, .
\end{equation}
Here, $\varphi_a$, $a = 1, \ldots , N$, is the order parameter ($\varphi^2 = \varphi_a \varphi_a$) and $\varepsilon$ the conserved density, while $\Omega$ and $\Omega_{\varepsilon}$ denote the relaxation and diffusion rate, respectively. 

The functional renormalization group is formulated in terms of a flow equation for the scale-dependent effective action $\Gamma_{k}$, which is the generating functional of one-particle irreducible correlation functions~\cite{Wetterich:1992yh}. In our case it depends on the field expectation values $\phi_{a} = \langle \varphi_{a} \rangle$, $\mathcal{E} = \langle \varepsilon \rangle$, as well as their corresponding response fields $\tilde{\phi}_{a}$ and $\tilde{\mathcal{E}}$ \cite{Martin:1973zz}. The exact flow equation in Fourier space is given by~\cite{Wetterich:1992yh}
\begin{equation}
\frac{\partial \Gamma_{k}}{\partial s} = \frac{1}{2} \Tr\! \int \!\! \frac{d^{d}q}{(2\pi)^{d}} \frac{d\omega}{2\pi}
\, \frac{\partial R_{k}}{\partial s} (q,\omega) \left( \Gamma_{k}^{(2)} + R_{k} \right)^{-1}\!\! (q,\omega) \, ,
\label{Eq:FlowEquation}
\end{equation}
where the logarithmic scale derivative is written in terms of $s = \ln (k/\Lambda)$ with some ultraviolet reference scale $\Lambda$, and the trace denotes a summation over fields. The second functional derivative of the scale-dependent effective action is given by $\Gamma_{k} ^{(2)}(q,\omega) \equiv \delta^{2} \Gamma_{k}/\delta \chi^{T} (-q,-\omega) \delta \chi(q,\omega)$,
where $\chi$ denotes the complete field content of our model. The regulator function $R_{k}$ implements a masslike cutoff and regulates the infrared modes. For $R_{k} \simeq 0$ the full generating functional of the theory $\Gamma = \Gamma_{k\rightarrow 0}$ is obtained. We use the regulator function $R_{k}(q) = Z_{k} (k^{2} - q^{2}) \theta(k^{2} - q^{2})$ for spatial momenta~\cite{Litim:2000ci}, which allows us to obtain fully analytic expressions for the nonperturbative $\beta$ functions.

To approximately solve the renormalization group equation \eqref{Eq:FlowEquation}, we consider a truncation for $\Gamma_k$ to leading order in the derivative expansion,
\begin{eqnarray}
\Gamma_{k} &=& 
\int\! d^{d}x \, dt \left\{ \tilde{\phi}_{a} \left( \Omega_{k}^{-1} \frac{\partial}{\partial t} - Z_{k} \nabla^{2} + \bar{\gamma}_{k} \mathcal{E} \right) \phi_{a} + \tilde{\phi}_{a} \frac{\partial U_{k}}{\partial \phi_{a}}  \right. \nonumber\\
&& -\: \Omega_{k}^{-1} \tilde{\phi}^{2} + \tilde{\mathcal{E}} \left( \frac{\partial}{\partial t} - Z_{\mathcal{E},k} \nabla^{2}  \right) \mathcal{E} - \frac{\bar{\gamma}_{k}}{2} \tilde{\mathcal{E}} \nabla^{2} \phi^{2} \nonumber\\
&& +\: \tilde{\mathcal{E}} \nabla^{2} \tilde{\mathcal{E}} \bigg\} ~. 
\label{Eq:AnsatzEffectiveAction}
\end{eqnarray}
This defines the effective theory at the scale $k$ in terms of the scale-dependent kinetic coefficient $\Omega_{k}$, the wave function renormalizations $Z_{k}$ and $Z_{\mathcal{E},k}$ in the two sectors, the derivative of the effective potential $\partial U_{k} / \partial \phi_{a}$, and the coupling $\bar{\gamma}_{k}$ between the sectors. The dynamic coefficient in the $\mathcal{E}$ sector is not renormalized, and we have set $\Omega_{\mathcal{E},k} = 1$ since only the ratio of the kinetic coefficients in the two sectors is relevant for the critical dynamics. Inserting Eq.\ \eqref{Eq:AnsatzEffectiveAction} into Eq.\ \eqref{Eq:FlowEquation}, we obtain the flow equations for the scale-dependent parameters.

To investigate the critical properties, we introduce the dimensionless renormalized field squared $\rho = k^{2-d} Z \,\phi^{2}/2$ and potential $u (\rho) = k^{-d} U (\rho)$, where we drop the labels referencing the scale $k$ in order to ease the notation. 
The dimensionless renormalized coupling between the sectors is given by $\gamma = k^{d/2 - 2} Z^{-1} Z_{\mathcal{E}}^{-1/2} \bar{\gamma}$. To characterize the behavior of the renormalized kinetic coefficient $\Omega^{-1} Z^{-1} Z_{\mathcal{E}}$, it is convenient to introduce the kinetic parameter $\kappa = 1/(1 + \Omega^{-1} Z^{-1} Z_{\mathcal{E}})$.
In the scaling regime we need the flow equations for the case of a nonvanishing rescaled field expectation value or potential minimum, $\rho_{0} \neq 0$, defined by $u'(\rho_{0}) = 0$ with $u'\equiv \partial u/\partial \rho$. 
At a fixed point, $\rho_0$ is constant and $\lim_{k\rightarrow 0} k^{d-2} \rho_0/Z$ denotes the order parameter~\cite{Berges:2000ew}. Using a polynomial expansion for the potential to fourth order in the fields around $\rho_{0}$, we obtain the flow equations for $\rho_{0}$ and the effective coupling $\lambda = u'' (\rho_{0}) - \gamma^{2}$:
\begin{eqnarray}
\frac{\partial \rho_{0}}{\partial s}  &=& (2 - d - \eta) \rho_{0} \nonumber\\
&& +\: 2 v_{d} \left\{ (N-1) l_{1}(0;\eta) + 3 l_{1}( 2 \rho_{0} \lambda;\eta) \right\}, \label{eq:BrokenPhaseOrderParameter}\\
\frac{\partial\lambda}{\partial s}  &=& (d - 4 + 2 \eta) \lambda \nonumber\\
&& +\: 2 v_{d} \lambda^{2} \left\{ (N-1) l_{2}(0;\eta) + 9 l_{2}( 2 \rho_{0} \lambda;\eta) \right\} .\label{eq:BrokenPhaseLambda}
\end{eqnarray}
Here $v_{d} = \left( 2^{d+1} \pi^{d/2} \Gamma\left( d / 2 \right) \right)^{-1}$ and the scalar anomalous dimension is defined as $\eta = - \partial \ln Z/\partial s$. The functions
$l_{n}(w;\eta) = (2n/d) \left( 1 - \eta/(2 + d)\right) (1+w)^{-n-1}$
parametrize the integral appearing from Eq.\ \eqref{Eq:FlowEquation} for the potential flow and describe the net decoupling of heavy modes~\cite{Berges:2000ew}. The flow equation for the coupling $\gamma$ reads
\begin{eqnarray}
\frac{\partial \gamma}{\partial s} &=& \left(d/2 - 2 + \eta + \eta_{\mathcal{E}}/2\right) \gamma \nonumber\\ 
&& +\: 2 v_{d} \gamma ( \lambda + \gamma^{2} ) \left\{ (N-1) l_{2}( 0 ; \eta) 
+ 3 l_{2}( 2 \rho_{0} \lambda ; \eta) \right\} , \nonumber\\
\label{eq:BrokenPhaseGamma}
\end{eqnarray}
which has an explicit dependence on the anomalous dimension $\eta_{\mathcal{E}} = - \partial \ln Z_{\mathcal{E}} / \partial s$. The scale dependence of the kinetic parameter takes the form
\begin{equation}
\frac{\partial \kappa}{\partial s} = \kappa (1- \kappa) \left\{ \eta_{\Omega}(\kappa) - \eta + \eta_{\mathcal{E}} \right\} ~,
\label{eq:RatioKineticCoefficients}
\end{equation}
which depends also on the scaling contribution to the renormalized kinetic coefficient, $\eta_{\Omega} = -\partial \ln \Omega^{-1}/\partial s$. The anomalous dimensions are given by
\begin{eqnarray}
&& \hspace{-20pt} \eta = 16 \frac{v_{d}}{d} \rho_{0} \lambda^{2} m_{2,2}( 0,  2 \rho_{0} \lambda ; \eta) 
~, \label{eq:eta}\\
&& \hspace{-20pt} \eta_{\mathcal{E}} = - 2 v_{d} \gamma^{2} \left\{ ( N-1 ) l_{2}( 0 ; \eta) + l_{2}( 2 \rho_{0} \lambda ; \eta) \right\} 
~, \label{eq:etaE}
\end{eqnarray}
\begin{eqnarray}
\eta_{\Omega} &=& \frac{2 v_{d}}{\rho_{0}} \bigg\{ l_{1}(0;\eta) + l_{1}(2 \rho_{0} \lambda; \eta) \nonumber\\ 
&& -\: 2 \, h_{1} \big( ( \lambda + \gamma^{2} ) \rho_{0} , \gamma^{2} \rho_{0} (1 - \kappa)/\kappa, (1 - \kappa)/\kappa;\eta \big) \bigg\} ~. \nonumber\\
\label{eq:AnomalousDimensions}
\end{eqnarray}
Here $m_{2,2}(w_{1},w_{2};\eta) = (1+w_{1})^{-2} (1+w_{2})^{-2}$ is $\eta$ independent in our case and $h_{1}$ is a similar threshold function. In the limit $\kappa \rightarrow 1$, we have $h_{1}(w,0,0;\eta) = l_{1}(w;\eta)$, and for general $\kappa$ the expression can be given in terms of hypergeometric functions.
\begin{figure}[!t]
\centering
\includegraphics[width=0.48\textwidth]{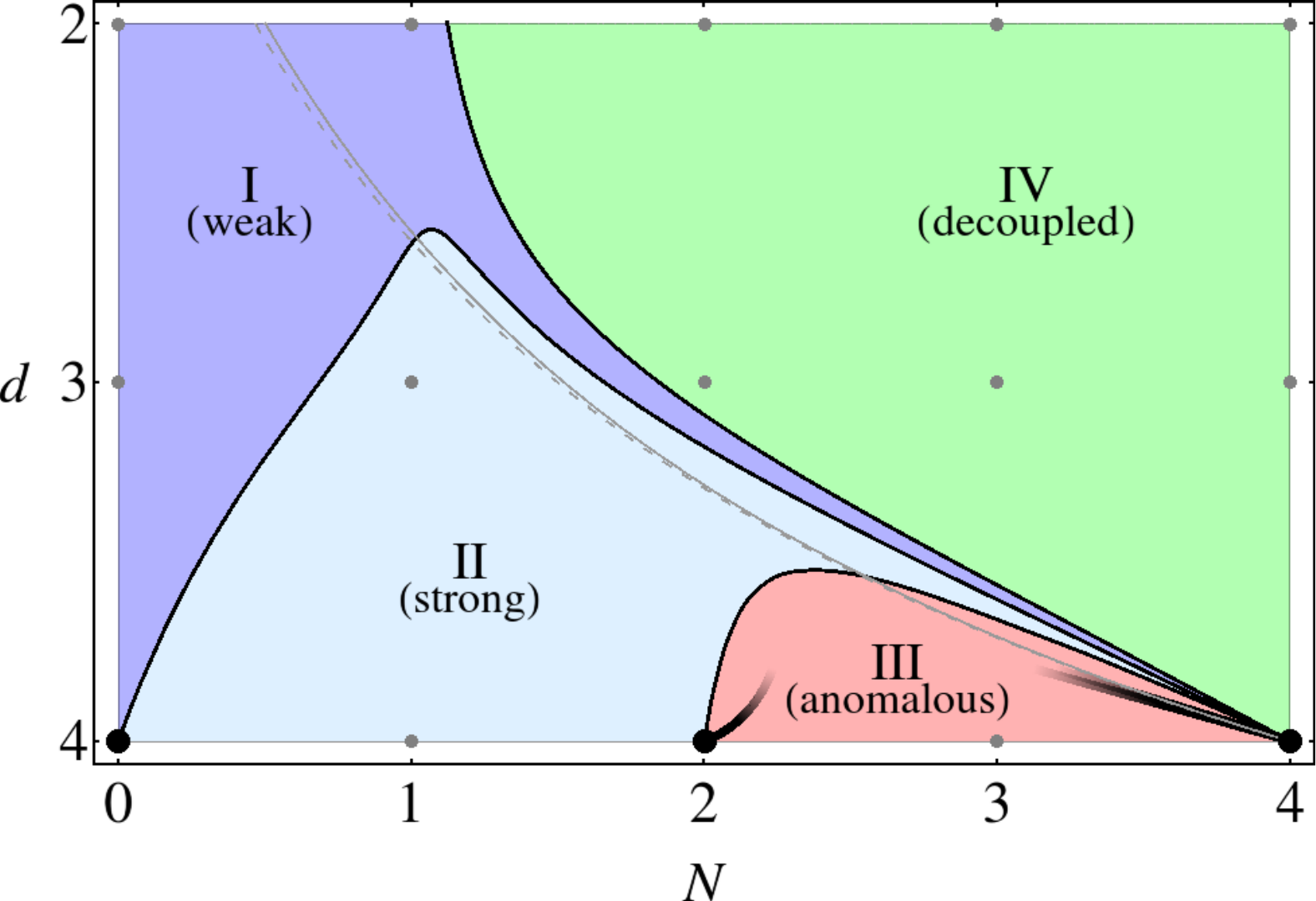}
\caption{\label{Fig:PhaseDiagram}(Color online) Phase diagram for Model C as a function of dimension $d$ and the number of field components $N$ from the functional renormalization group. For comparison to the $\epsilon$ expansion, thick lines near $d = 4$ denote $\mathcal{O}(\epsilon)$ results \cite{Brezin:1975tq}; thin/dashed lines denote the $\mathcal{O}(\epsilon^{2})$ results according to \mbox{Ref.\ \onlinecite{Folk:2003aa,*Folk:2004aa}}.
}
\end{figure}

Equations \eqref{eq:BrokenPhaseOrderParameter} -- \eqref{eq:AnomalousDimensions} constitute the full set of flow equations for this model, whose fixed point solutions with vanishing scale derivative are computed numerically. At a fixed point $Z \sim k^{-\eta}$, $Z_{\mathcal{E}} \sim k^{-\eta_{\mathcal{E}}}$, and $\Omega^{-1} \sim k^{-\eta_{\Omega}}$ assume their scaling form while the anomalous dimensions $\eta$, $\eta_{\mathcal{E}}$, and $\eta_{\Omega}$ take on their scale-independent critical values. From these the dynamic critical exponents are defined as $z = 2 - \eta + \eta_{\Omega}$ and $z_{\mathcal{E}} = 2 - \eta_{\mathcal{E}}$, respectively.
The static critical behavior is encoded in the flow equations \eqref{eq:BrokenPhaseOrderParameter} and \eqref{eq:BrokenPhaseLambda} characterizing the potential flow with the anomalous dimension [Eq.\ \eqref{eq:eta}]. They form a closed set of equations and only depend on $N$ and $d$ for the $O(N)$ symmetric potential, which reflects the fact that the static universality class does not depend on the dynamic properties. 

In addition to the static properties, the dynamic universality class is further characterized in terms of the fixed point values of $\gamma$ and $\kappa$ along with the scaling exponents $z$ and $z_{\mathcal{E}}$. 
Our results for the ($N,d$) phase diagram for the dynamic critical behavior are shown in Fig.~\ref{Fig:PhaseDiagram}, where we find the following distinct scaling regions:

I. \emph{Weak scaling region:} For $\kappa = 0$ and $\gamma \neq 0$ at the fixed point, we obtain two 
independent dynamic scaling exponents $z$ and $z_{\mathcal{E}}$, where $z > z_{\mathcal{E}}$. Since the ratio of the renormalized relaxation rate and the diffusion rate vanishes, the 
order parameter relaxes only asymptotically compared to the diffusion timescale in this regime.

II. \emph{Strong scaling region:} For $0 < \kappa < 1$ and $\gamma \neq 0$, we find from Eq.\ \eqref{eq:RatioKineticCoefficients} with $\partial \kappa/\partial s = 0$ at the fixed point that $\eta_{\Omega} - \eta + \eta_{\mathcal{E}} = 0$. This leads to a \emph{locking of the dynamic critical exponents} in both sectors, with $z = 2 - \eta_{\mathcal{E}} = z_{\mathcal{E}}$.
This strong scaling holds when the fluctuations of the conserved density dictate the dynamic critical scaling for the order parameter. It is this region that is commonly referred to as Model C.

III. \emph{Anomalous diffusion region:} For critical $\kappa = 1$ and $\gamma \neq 0$, we find another weak scaling solution with independent values for the scaling exponents, i.e., $z < z_{\mathcal{E}}$, in contrast to region I. Here, the ratio of the renormalized diffusion rate and the relaxation rate vanishes, which describes the peculiar situation of a purely diffusive process in the presence of a homogeneous order-parameter field.   

IV. \emph{Decoupled scaling region:} If the two sectors decouple, i.e., $\gamma = 0$, then $\eta_{\mathcal{E}} = 0$ according to Eq.\ \eqref{eq:etaE}. In this case, the conserved density displays dimensional scaling with $z_{\mathcal{E}} = 2$. In this region, the physical field shows a dynamic critical scaling with leading exponent $z$ in the \mbox{Model A} universality class. However, there can be nontrivial subleading corrections to the dynamic scaling even if the mode coupling is zero~\cite{Brezin:1975tq}.
\begin{figure}[!t]
\centering
\includegraphics[width=0.48\textwidth]{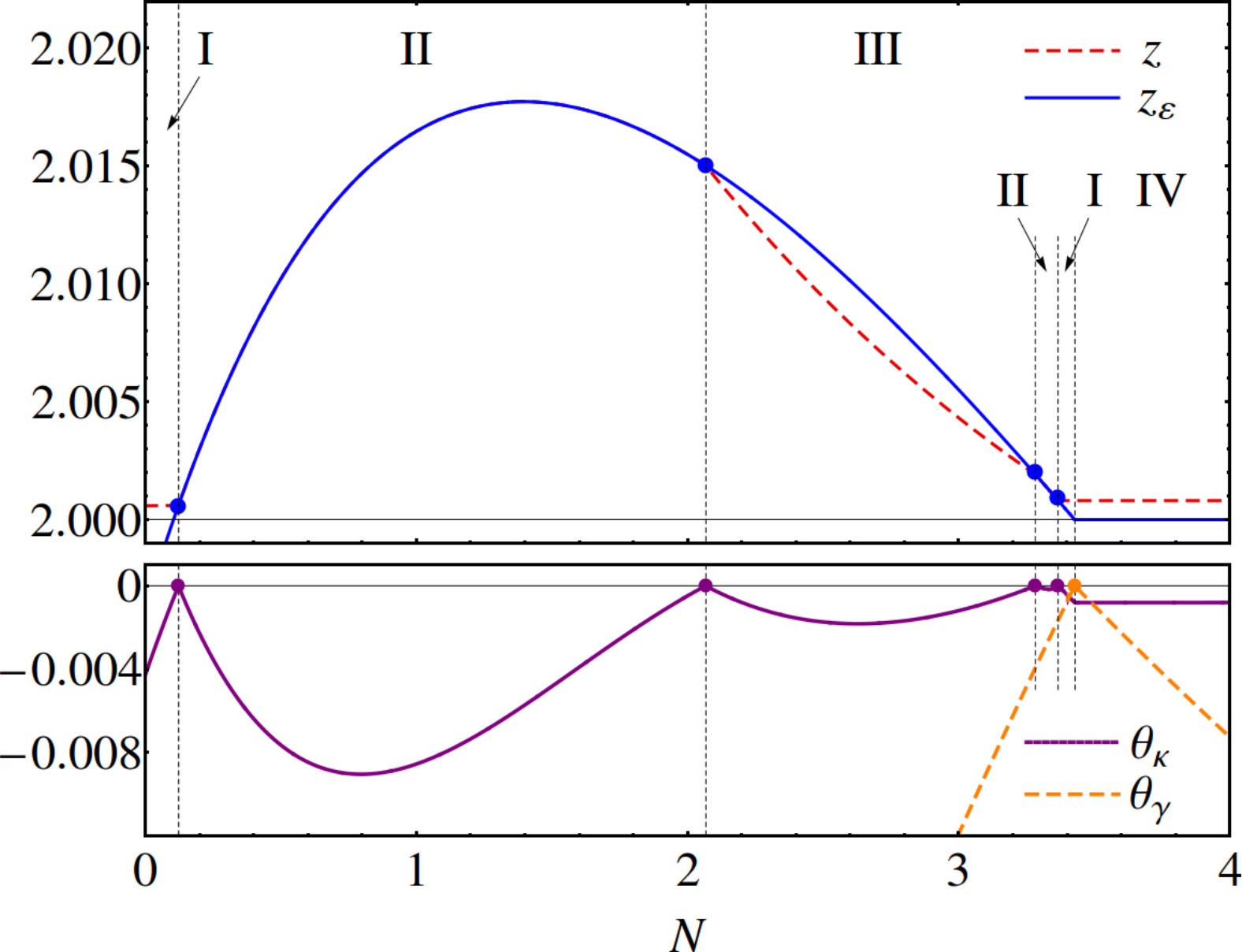}
\caption{\label{Fig:Exponents}(Color online) (\textbf{Top}) Dynamic critical exponents $z$ and $z_{\mathcal{E}}$ as a function of $N$ at fixed dimension $d = 3.75$. The different regions I -- IV are clearly visible, along with the locking phenomenon ($z=z_{\mathcal{E}}$) in II and the distinct values in the independent scaling regimes I and III. (\textbf{Bottom}) The subleading exponents $\theta_{\kappa}$ and $\theta_{\gamma}$ indicate the stability of the fixed point solutions.}
\end{figure}

\begin{figure}[!t]
\centering
\includegraphics[width=0.48\textwidth]{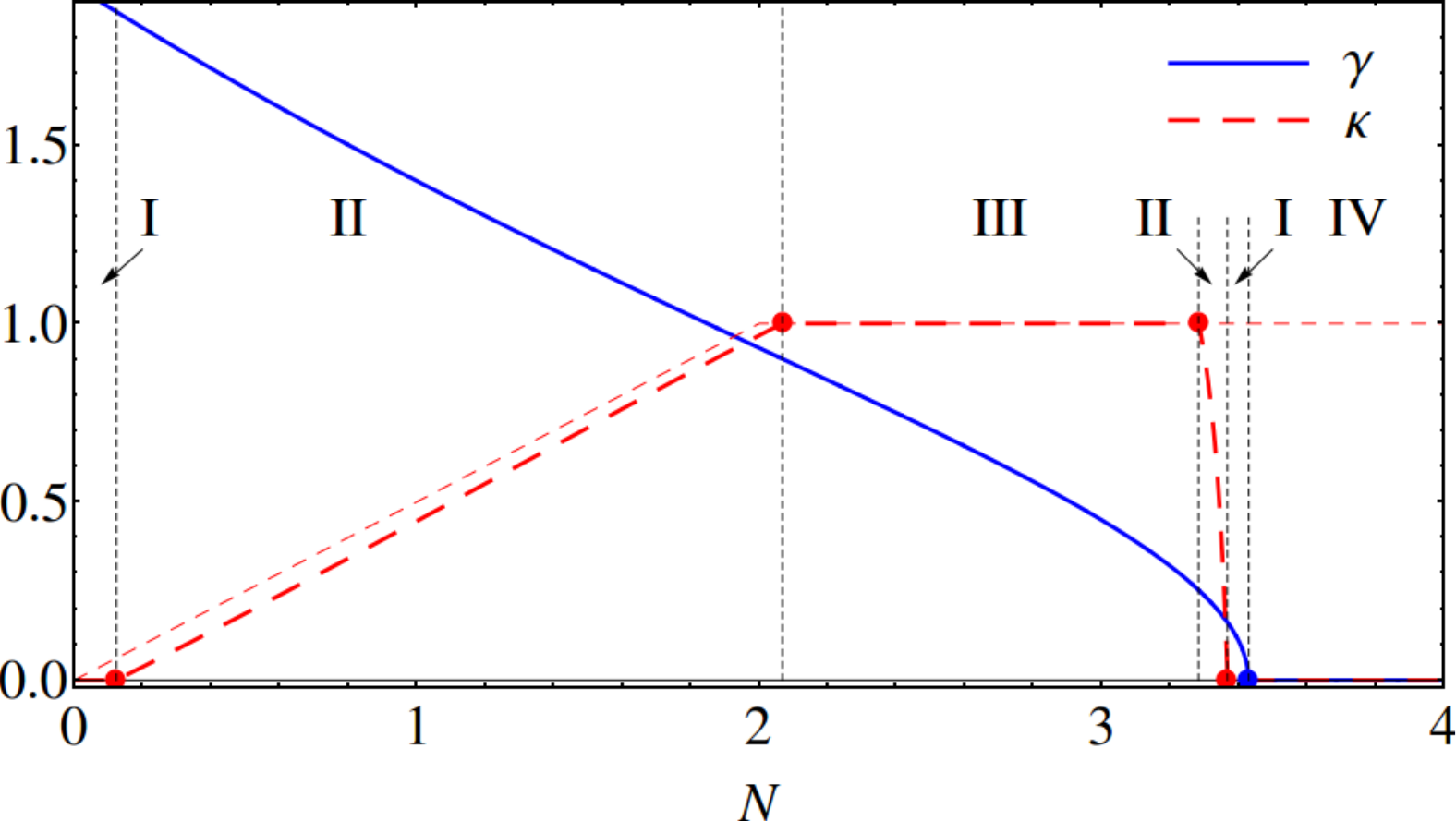}
\caption{\label{Fig:Couplings}(Color online) Fixed point values for the mode coupling $\gamma$ and kinetic parameter $\kappa$ as a function $N$ at $d = 3.75$. The asymptotic result for $\kappa$ in the limit $d \rightarrow 4^{-}$ is shown for comparison. In that case region III extends over the range $2 < N < 4$.}
\end{figure}

The subleading exponents also give information about the stability of the fixed point solutions. The eigenvalues of the stability matrix $\partial \beta_{i} / \partial g_{j}$, which we write in terms of the generalized couplings $g_{i} \in \{ \lambda , \ldots \}$ with $\beta_\lambda \equiv \partial \lambda/\partial s$ etc., define the critical exponents. There are two independent static exponents, and our results for the correlation length exponent $\nu$ and for $\eta$ accurately agree with those documented for functional renormalization group studies on the static universality class at this truncation level~\cite{Berges:2000ew}. For our analysis it is important that we can extract subleading exponents from the stability matrix. The characteristic behavior of the eigenvalues $\theta_{\kappa} = \partial \beta_{\kappa} / \partial \kappa$ and $\theta_{\gamma} = \partial \beta_{\gamma} / \partial \gamma$ is exemplified in Fig.\ \ref{Fig:Exponents}. These eigenvalues are negative except at the boundaries between the phases I -- IV, where the different fixed point solutions exchange their stability properties and either $\theta_{\kappa}$ or $\theta_{\gamma}$ changes sign if evaluated beyond the stable regime. Figure \ref{Fig:Couplings} shows the corresponding fixed point values of $\gamma$ and $\kappa$ that define the scaling regions I -- IV.

Using more sophisticated truncations to higher orders in the derivative expansion and extending the basis of field operators would be required to obtain an error estimate for our results, which goes beyond this work. However, we observe that our phase diagram is compatible with known data for both critical dynamics and statics. Model C is special compared to the other relaxational universality classes, as it relates the dynamic to the static scaling properties close to criticality. In particular, at a fixed point the two-point correlation function for the conserved density assumes a scaling form $\langle \varepsilon \varepsilon \rangle_{k} \sim k^{\eta_{\mathcal{E}}}$, and we have $\eta_{\mathcal{E}} = - \alpha / \nu$ when $\alpha > 0$, while such a relation is absent for $\alpha < 0$, when the coupling to the conserved density renormalizes to zero ($\gamma = 0$) \cite{Brezin:1975tq}. Since the boundary between regions I and IV is characterized by the vanishing of the coupling $\gamma$, one may deduce its location from static equilibrium properties where $\alpha = 0$. Even for the two-dimensional Ising model, where $\alpha = 0$ is known exactly from the Onsager solution, our result for the phase boundary in Fig.~\ref{Fig:PhaseDiagram} still occurs remarkably close to the exact result in comparison to the $\epsilon$ expansion. Furthermore, in the strong scaling region, we can directly compare our results for the dynamic critical exponent to the value obtained from the corresponding scaling relation $z = 2 + \alpha/\nu$. Using most accurate high-temperature expansion data for $N=1$ in $d=3$ from Ref. \onlinecite{Pelissetto:2000ek}, we obtain from this relation $z = 2.176(3)$ and compare to our result $z = 2.059$, which is reasonable for a lowest-order derivative expansion in the presence of sizable anomalous dimensions. 

In the weak and decoupled scaling regions (I and IV), where $\kappa = 0$, the leading dynamic scaling behavior of the physical field features a dynamic exponent in the universality class of \mbox{Model A}. In fact, we may compare the contributions to $z - 2$ directly to the flow equations for \mbox{Model A} \cite{Canet:2006xu} and find $\eta_{\Omega} - \eta = c \eta$ in this region using the standard notation \cite{Hohenberg:1977ym}. Knowledge about the values of $c \eta$ allows us to deduce the shape of the transition between the weak and strong scaling regions (I and II), which is characterized by the locking of dynamic critical exponents $z = z_{\mathcal{E}}$. In particular, the boundary is defined by the relation $\alpha/\nu = c \eta$. Using available data on the quantity $c \eta$ from the critical dynamics of \mbox{Model A} \cite{Bausch:1981zz,Albano:2011aa} and the static critical exponents $\alpha$ and $\nu$ \cite{Pelissetto:2000ek}, we find that the phase boundary for $N = 1$ should pass between $2 < d < 3$, which is in very good agreement with our results. However, in the limit $N \rightarrow 0$ the situation is less clear -- this applies to the bending down of the boundary between regions I and II. Data from self-avoiding random walk (SAW) models \cite{BenAvraham:1982aa} for the case $N = 0$ and field-theoretic results \cite{Pelissetto:2000ek} indicate that $\alpha/\nu$ is positive between the upper and lower critical dimension $1 < d < 4$, while we find a small negative contribution to the dynamic critical exponent of the conserved density, i.e., $z_{\mathcal{E}} - 2 < 0$ (as seen also in Fig.~\ref{Fig:Exponents}). The dynamic critical exponent $z$, however, receives a positive contribution in this regime and is compatible with a lower bound derived for the relaxational models \cite{Katzav:2011aa}. Monte Carlo simulations for SAW models in fractal dimension within the Model C dynamic universality class could clarify the situation and pin down the structure of the phase diagram in the $N \rightarrow 0$ region.

It would be striking if one could establish the scaling properties of region III experimentally. This region describes a diffusion process in the presence of a homogeneous scalar field configuration. Nevertheless, fluctuations of the order parameter are important and the nonzero coupling $\gamma \neq 0$ strongly affects the scaling properties of the conserved density, i.e., $z_{\mathcal{E}} = 2 - \eta_{\mathcal{E}} > 2$, which leads to \emph{subdiffusion} without disorder. It would be interesting to see if this region of the phase diagram is accessible with Monte Carlo simulations for fractal dimensions $3 < d < 4$ if the real-time dynamic critical behavior is identified with the dynamic properties of the Monte Carlo sampling process \cite{Muller-Krumbhaar:1973,Li:1994he}. Also, diluted scalar models might yield an indication for such a phase.

\begin{acknowledgments}

We thank L.~Can\'et, J.\,M.~Pawlowski, M.\,M.~Scherer, U.\,C.~T\"auber, and N.~Wschebor for discussions. D.\,M. is supported by the Deutsche Forschungsgemeinschaft within the \mbox{SFB 634}. L.\,F.\,P.\ is supported by a fellowship of the Alexander von Humboldt foundation.

\end{acknowledgments}

\bibliography{references}

\end{document}